\def\x{\vec x_1} 
\def\p{\vec p_1}

\def\rb{{\vec{x}}}
\def\pb{{\vec{p}}}

\def\pp{\vec{p}\,'}

\def\be{\begin{equation}}
\def\ee{\end{equation}}
\def\bea{\begin{eqnarray}}
\def\eea{\end{eqnarray}}
\documentstyle[12pt,dina4,psfig,preprint,aps]{revtex}
\begin{document}
\draft  
\title{Can one extract source radii from transport theories?}


\author{J. Aichelin}
\address{SUBATECH \\
Laboratoire de Physique Subatomique et des Technologies Associ\'ees \\
UMR Universit\`e de Nantes, IN2P3/CNRS, Ecole des Mines de Nantes\\
4, rue Alfred Kastler 
F-44070 Nantes Cedex 03, France.}
%
\maketitle
\widetext
\begin{quote}
\begin{abstract}
To know the space time evolution of a heavy ion reaction is of great 
interest, especially in cases where the measured spectra do not
allow to ascertain the underlying reaction mechanism. In recent times 
it became popular to believe that the comparison of 
Hanbury-Brown Twiss correlation functions obtained from classical or 
semiclassical transport theories, 
like Boltzmann Uehling Uhlenbeck (BUU), Quantum Molecular Dynamics (QMD), 
VENUS, RQMD or ARC, with experiments may provide this insight. 
It is the purpose of this article to show that this conjecture 
encounters serious 
problems. The models which are suited to be compared with the 
experiments at CERN and Brookhaven are not able to predict a
correlation function. Any agreement with existing data has to be
considered as accidental. The models suited for lower energies can
in principle predict correlation functions. The systematic error
may be too large to be of use as far as quantitative conclusions
are concerned.    
\end{abstract}

\end{quote}
\narrowtext 
\pagebreak 
\psdraft 

\section{Introduction}
It is a common problem in heavy ion reactions between 25 MeV/N and 200 GeV/N
that the single particle spectra do not allow to ascertain the underlying
reaction mechanism. To mention only two examples: At low energies despite of 
many years of efforts the fragmentation of nuclei into many intermediate
mass fragments remains still a process whose origin is heavily
debated. At high energies it turned out to be very difficult to 
rule out a hadronic scenario which may produce the same spectra
as those proposed as a signal for the creation of a quark gluon plasma.

In such a situation a search for experimental information beyond the
single particle spectra is obvious. Most valuable would be an information
on the spatial structure of the reaction. It would allow to calculate
key quantities like densities or energy densities. This information is, 
however, hard to obtain. 

The only promising method proposed up to now is based on the interferometry 
of identical particles. The interference of the amplitudes of two
indistinguishable processes gives rise to a correlation function which
in principle allows to extract the radius of the emitting source. This
approach has been  very successfully applied by Hanbury-Brown and Twiss 
\cite{hbt} in astronomy to determine the angular radius 
of stars by measuring the spatial correlations between two photons.
Later Goldhaber\cite{gol} and Kopylov and Podgoretsky 
\cite{Podgor72} advanced its application in particle physics by showing that
measurable momentum space correlations may contain information on the 
size of the emitting source.

In the ideal case of a large, randomly emitting source of known shape
this method is indeed very powerful and the experimental results can be 
directly related to the source radius of the emitting object. 
In particle and heavy ion physics the situation is, however, much more 
difficult. 
There we encounter quite a number of problems. The size of the emitting 
sources is of the order of the radius of a nucleus and therefore not small
as compared to the size of the wave function of the emitted particles.
This renders some approximations impossible.
The signal may be distorted by final state interactions between the 
emitted particles or by to the
long range Coulomb force of the source. The emission time point cannot
be defined unambiguously. The decay of resonances into identical 
particles or correlations between the momenta of the emitted particles 
and the coordinates of the emission point may pretend a wrong size
of the source. For a discussion of these problems we refer to ref. \cite{zaj}
,\cite{gyu}. Recently is has been discussed that the HBT correlation
function for an expanding source, as encountered frequently in heavy ion 
reactions, yields a much more difficult
relation between the space time structure of the emitting source and
the correlation function as that for a static source\cite{hei}.

Due to these problems the measured correlation function 
in heavy ion collisions cannot be
directly related to the parameters of the emitting source even if its
form were known. In this situation there are two possibilities. Either
one {\it assumes} the form of the source and uses the measured correlation
function to fix the source parameters. Unfortunately this procedure
makes these parameters model dependent. Hence
they cannot be used for more than a comparison between different
experiments and yield little information on the actual source properties.
Or one tries to describe the reaction in its entity. This
turns out to be a quite complicated procedure but became very popular
recently. In this approach one follows the time evolution of the system
with help of one of the standard transport models. Unfortunately none of 
them propagates (anti)symmetrized wave functions but at most a direct
product wave function. Since the HBT effect is based on the
(anti)symmetrization of the wave function of identical particles the 
transport model themselves cannot predict the correlation function.
Rather one assumes that each particle "freezes out" at some time point. The
freeze out time is different for each particle. At high energies it is
assumed that the freeze out time is that time at which the particle
encounters its last collision with another particle of the system.
At low energies, where potential interactions are important as well, 
the freeze out time cannot be unambiguously defined. 
The freeze out times as well as the particle momenta and positions form then the
input for the subsequent calculation of the Hanbury-Brown and Twiss 
(HBT) correlation function \cite{led} which is then compared with experiment. 

Agreement is usually interpreted as a sign that the underlying transport
model gives a realistic space time evolution of the different particles. 
Since these transport codes provide not only the momentum space coordinates
of the particles but also that of the coordinate space they can then be
used to calculate the time evolution of key quantities like the 
energy density or the density. 

The weak point in this procedure is the transition between
the transport model and the subsequent program which calculates the
correlation function. Does the transport model provide the correct
time evolution of those quantities which are essential for the
calculation of the correlation function? 

It is the purpose of this article to show that this is hardly the case. 
In order to understand the reason one has to understand in detail the 
derivation
of the different transport models from the fundamental quantal 
equations as well as the derivation of the equation which is employed to
determine the correlation function. We will perform this investigation for the
three types of present day simulation programs: The Quantum Molecular
Dynamics approach (QMD)\cite{aichelin91}, 
BUU type models like 
Boltzmann \"Uhling Uhlenbeck (BUU) \cite{ai85a}-\cite{cas90}, 
Vlasov \"Uhling Uhlenbeck (VUU) 
\cite{st86} or Landau Vlasov (LV) \cite{greg87} and
cascade models. The later class includes also the high energy simulation
programs like VENUS \cite{wer}, RQMD \cite{rqmd} and ARC \cite{arc}.

This problem is independent of the relativistic or nonrelativistic 
nature of the approach. It also does neither depend on the time
between the emissions of the two identical particles nor on the
presence of resonances. It is also independent of a possible final state
interaction between the particles which is therefore omitted.
The common demand on all transport programs is that 
an emission time point can be defined. Essential is the information
the programs provide at that time point. This information is quite different
for the three types of programs mentioned above and hence the systematic
errors are specific to each of the different transport models.
In two cases (QMD and BUU) this procedure implies systematic errors
which question the usefulness of the approach for its original purpose: The
discrimination between different reaction mechanisms which yield
the same single particle spectra. For the high energy simulation 
programs the correlation function is completely artificial. 

For clarity we limit our formalism to the simplest
form possible by assuming that we are dealing with two bosons which are
simultaneously emitted and can be treated nonrelativistically. For
this simple case the formalism is very transparent. More realistic
but also more complicated scenarios \cite{hei} may add additional problems
but do not overcome the problems discussed here.

\section{The correlation function}
We start with the derivation of the correlation function 
which relates the freeze out points with the measurable two 
body correlation function. We assume that
a source, which is considered as classical, emits simultaneously
two identical bosons with momenta $\vec p_1$ and $\vec p_2$. The 
differential two body probability W reads then as follows \cite{Pratt84}  
($\hbar,c =1$): 

\begin{equation}
\label{1}
{d^2W \over d\vec p_1 d\vec p_2} = |T_S(\vec p_1, \vec p_2,
\alpha ,\beta)|^2
\end{equation}
\noindent
$\alpha $ and $\beta $ characterize the emitting source.
The (anti)symmetrized production amplitude $T_S$ is given by

\begin{equation}
\label{2}
T_S(\vec p_1,\vec p_2,\alpha ,\beta) = {1\over\sqrt{2}}
(T (\vec p_1, \vec p_2,\alpha ,\beta) \pm T(\vec p_2,
\vec p_1,\alpha ,\beta)).
\end{equation}
\noindent
where $T(\vec p_1, \vec p_2,\alpha ,\beta)$ 
is the Fourier transform of the wave function  

\begin{equation}
\label{3}
T(\vec p_1, \vec p_2,\alpha ,\beta) = \int{d^3x_1d^3x_2 \over (2\pi)^3} 
e^{-i(\vec p_1\vec x_1+\vec p_2\vec x_2)}<\vec x_1,\vec x_2|
\psi(\alpha ,\beta)>.
\end{equation}
Introducing the Wigner density of the two body density matrix $\rho_2 =
|\psi_2><\psi_2|$
\begin{equation}
\label{4}
D(\vec x_1,\vec p_1,\vec x_2,\vec p_2) = {1\over (2\pi)^6}\int 
\prod_{i=1,2}d^3y_i e^{i\vec p_i\vec y_i}
<\vec x_1-\vec y_1/2,\vec x_2-\vec y_2/2|\psi_2><\psi_2|\vec x_1+\vec y_1/2,
\vec x_2+\vec y_2/2>
\end{equation}
we can express the probability as a function of the two particle
Wigner density \cite{gyu}
\begin{equation}
\label{1}
{d^2W \over d\vec p_1 d\vec p_2} = \int d^3x_1d^3x_2[
D(\vec x_1,\vec p_1,\vec x_2,\vec p_2)\pm D(\vec x_1,{\vec p_1+\vec p_2\over 2},
\vec x_2,{\vec p_1+\vec p_2\over 2})cos(\vec p_1-\vec p_2)(\vec x_1-\vec x_2)].
\end{equation}
Hence for the calculation of this probability the transport theories 
have to provide the two body Wigner density. 
Unfortunately most of them do not permit to calculate this quantity.
Therefore one has introduced an approximation, called smoothness assumption
(SA) :
\be
D(\vec x_1,{\vec p_1+\vec p_2\over 2},\vec x_2,{\vec p_1+\vec p_2\over 2}) 
\approx 
D(\vec x_1,\vec p_1,\vec x_2,\vec p_2).
\ee
We will discuss the limits of its validity which turns out to be crucial
in the course of the article.
Employing the smoothness assumption the two body probability reads as
\be
{d^2W^{SA} \over d\vec p_1 d\vec p_2} = \int d^3x_1d^3x_2
D(\vec x_1,\vec p_1,\vec x_2,\vec p_2)
(1 \pm cos(\vec p_1-\vec p_2)(\vec x_1-\vec x_2)).
\ee
This is the standard expression for the two particle probability 
employed in numerous publications to relate the measured cross section
with the radius of the emitting source.

Depending on the available information, for actual calculations one may have 
to employ further approximations. For BUU type equations, which propagate the 
one particle Wigner density only, one assumes that the correlations 
between particles are negligible
\be
D(\vec x_1,\vec p_1,\vec x_2,\vec p_2)\approx
D(\vec x_1,\vec p_1)\cdot D(\vec x_2,\vec p_2)
\ee
whereas for classical cascade calculations one assumes that the quantal 
two body Wigner density can be replaced by the classical 2 body phase 
space density $F_{cl}$
\be
D(\vec x_1,\vec p_1,\vec x_2,\vec p_2)\approx
F_{cl}(\vec x_1,\vec p_1,\vec x_2,\vec p_2).
\ee
For a static source without any correlation between the emission point
and the momentum of the emitted particle the 
correlation function, the quantity one compares with experiment, is
independent of the center of center of mass motion of the emitted pair
and is given by 
\be
C(\vec p) ={\int {d^2 W \over d\vec p_1 d\vec p_2} d^3P \over
\int { dW \over d\vec p_1 }{dW \over d\vec p_2}d^3P} 
\ee 
where $\vec P = {(\vec p_1+\vec p_2)\over 2}$ is the center of mass momentum,
$\vec p = \vec p_1-\vec p_2$ is the relative momentum and $ 
{dW \over d\vec p_1 }$  is the one particle momentum distribution
\be
{dW \over d\vec p_1 }= \int d^3x_1 D(\vec x_1, \vec p_1).
\ee   
In the general case, where correlations are present, C depends on the
center of mass motion as well. 
As we will see the correlation function $C(\vec p)$ contains the desired
information about the spatial properties of the emitting source.

\section{Consequences of the smoothness assumption for cascade
calculations}
One of the classes of models employed to extract source radii by comparing
experimental results with model predictions are the so called cascade 
models. These are classical n-body models which solve the Hamilton
equations of a n-body system and are presently the only models available 
to simulate heavy ion reactions at CERN and Brookhaven energies. 
They include VENUS \cite{wer}, RQMD \cite{rqmd} (in its usually employed
cascade version) and ARC \cite{arc} 
as well as now less frequently employed programs for
heavy ion reactions at an energy of around 1 GeV/N \cite{cug}.
In these models the particles do not interact via potentials
but suffer two body collisions if they come sufficiently close in
coordinate space. In between the collisions the particles move on
straight lines. In the computer programs they are treated as classical
particles with a sharp momentum and a sharp position. 

One may ask how classical models can be employed to calculate a 
correlation function which is solely based on the interference of 
amplitudes and hence a genuine quantal effect. 
For an understanding we have to make a detour. In order to employ
eq. 5 we have to construct the Wigner density out of 
the classical two body phase space density. This is of course not
unique but the approach
\bea
F_{cl} &=& \delta(\x -\vec x_{\alpha})\delta(\p -\vec p_{\beta})
\delta(\rb_2 -\vec x_{\beta})\delta(\pb_2 -\vec p_{\beta}) \nonumber \\ 
&=& \lim_{C\to \infty,D\to 0} {C^3D^3\over \pi^6}
e^{-(\vec P -\vec K(t))^2C/4-(\vec X -\vec R(t))^24/D}
e^{-(\vec p -\vec k(t))^2C-(\vec x -\vec r(t))^2/D} \nonumber \\ 
&\equiv& D_{cl}(\vec x_1,\vec p_1,\vec x_2, \vec p_2)
\eea
serves our purpose. The expression in the last line will be 
considered as Wigner density.Here we have used the definitions 
\bea
\vec P = \vec p_1+\vec p_2&;\vec K = \vec p_\alpha+\vec p_\beta\nonumber \\  
\vec X = {\vec x_1+\vec x_2 \over 2}&;
\vec R = {\vec x_\alpha+\vec x_\beta \over 2}\nonumber \\  
\vec p = {\vec p_1-\vec p_2 \over 2}&;
\vec k = {\vec p_\alpha-\vec p_\beta \over 2}\nonumber \\  
\vec x = \vec x_1-\vec x_2&; \vec r  = \vec x_\alpha-\vec x_\beta.  
\eea 
The Wigner density $
D_{cl}(\vec x_1,{\vec p_1+\vec p_2\over 2},
\vec x_2,{\vec p_1+\vec p_2\over 2})$ is obtained by replacing $\vec p_1$ 
and $\vec p_2$ by ${\vec p_1+\vec p_2\over 2}$.

Please note that this Wigner density does not respect
the uncertainty relation. Inserting this expression in eqs. 5 and 7 and
performing the limit procedure we
obtain for the two particle correlator without smoothness assumption
\bea
{d^2W \over d\vec p_1 d\vec p_2}d^3P &=& 
\int d^3x_1d^3x_2d^3P[
D_{cl}(\vec x_1,\vec p_1,\vec x_2,\vec p_2)\pm D_{cl}(\vec x_1,{\vec p_1+\vec p_2\over 2},
\vec x_2,{\vec p_1+\vec p_2\over 2})cos(\vec p_1-\vec p_2)(\vec x_1-\vec x_2)]
\nonumber \\ 
&=& \delta(\vec p- \vec k(t)) \pm \delta(\vec k(t))cos(2 \vec p \vec r). 
\eea
This expression differs from 
\be
\int { dW \over d\vec p_1 }{dW \over d\vec p_2}d^3P
=\delta(\vec p- \vec k(t))
\ee
only for the case that the relative momentum of the emitted classical 
particles is zero what in practical terms never
happens. For all other cases we find
\be 
C(\vec p) = 1.
\ee
Applying the smoothness assumption we find, however, 
\be
\int{d^2W^{SA} \over d\vec p_1 d\vec p_2} d^3P
 = \delta(\vec p-\vec k(t))(1+cos 2\pb\vec r	)
\ee 
and hence
\be
C^{SA}(\vec p) =  1+cos 2\pb\vec r.
\ee
Thus we observe that here the smoothness assumption creates
correlations out of nothing. One faces the somewhat surprising result
that the correlation function and hence the extracted radii are artificial
and are only due to the differences between the approximate and 
the exact formula for the correlation function.  
Applying the correct formula the cascade calculations do
not yield any correlation function, as the exact result shows. 

The truth of this observation can even easily be verified without 
any calculation. If two
particles with a sharp momentum are emitted from two localized sources
one can measure the momentum sufficiently precise in order to 
identify the source from which each particle has been emitted
provided the both momenta are not identical.
Thus there are no alternative processes, hence no interference of their 
amplitudes and there is, as a consequence, no HBT correlation function.  
{\it This has unfortunately the consequence
that there is presently no microscopic model which may be used 
for the interpretation of the correlation data 
measured with ultrarelativistic heavy ion beams at CERN and AGS}.

\section{Quantum Molecular Dynamics (QMD)}
The discussion of models which provide sufficient information to
construct a correlation function we begin with the QMD approach because 
it is the only one which allows to calculate the 2 body Wigner density. 
Hence one can calculate ${d^2W \over d\vec p_1 d\vec p_2}$
(eq.5) without any approximation. One can furthermore
introduce the smoothness assumption and can calculate then 
${d^2W^{SA} \over d\vec p_1 d\vec p_2}$ applying eq. 7.
This may serve as a test for
the validity of this approximation in the situation of a heavy ion
reaction and hence for the judgement of the predictive power of the 
correlation function
calculated in the framework of the other models.

The  QMD model is a n body theory which simulates heavy ion reactions between
30 MeV/N and 2 GeV/N on an event by event basis. Each nucleon is represented
by a coherent state of the form  \be
\phi_\alpha (\p,t) = \left({\frac{L }{2\pi}}\right)^{3/4}\, e^{-(\p -
\vec p_\alpha(t))^2L/4} \,e^{-i\p \vec x_\alpha(t)}\,
e^{+i p_\alpha^2(t)t/2m} \ee
Thus the wave function has two time dependent parameters $x_\alpha, p_\alpha$%
, L is fixed. As we will see this wave function serves as a test wave function
for a variational principle. 
Hence it is an {\it input} of the calculation and not the
result of the solution of the Schr\"odinger equation. It
relies heavily on  intuition; other test wave functions may yield a different
time evolution of the system.
The total n body wave function is assumed to be the direct
product of n coherent states \be
\phi = \phi _\alpha (\x, \vec x_\alpha, \vec p_\alpha, t) 
\phi _\beta (\vec x_2,\vec x_\beta,\vec p_\beta, t)\cdots, \ee
thus antisymmetrization is neglected. The initial values of the parameters
are chosen in a way that the ensemble of $A_T$ + $A_P$ 
nucleons gives a proper density distribution as well as a proper momentum
distribution of the projectile and target nuclei. The time evolution of the
system is calculated by means of a generalized variational principle: 
We assume that $\vec p_\alpha$ and $\vec x_\alpha$contain  the essential
time dependence of the n-body wave function. The Lagrange function 
${\cal L}$ can then be written as a functional of these parameters
where H is the n - body Hamiltonian . 
 \be
{\cal L} = \left(\phi \left\vert i\hbar ({\frac{\partial }{\partial t}}
+{\frac{d\vec p_\alpha }{dt}}
{\frac{d}{d\vec p_\alpha}}+{\frac{d\vec x_\alpha }{dt}}
{\frac{d}{d\vec x_\alpha}}) - H \right\vert
\phi\right). \ee
The time evolution of the parameters is obtained by the
requirement that the action \be
S = \int\limits ^{t_2} _{t_1} {\cal L} [\phi, \phi^\ast] dt \ee
is stationary under the allowed variation of the wave function. 
For the wave function of eq. 20   the Lagrange 
function is given up to a constant by
\be
{\cal L} = \sum_\alpha(\vec p_{\alpha} \dot{\vec x_{\alpha}} -
{\vec p_{\alpha} \dot{\vec p_{\alpha}}t \over m} - 
{p_\alpha^2 \over 2m} - {1\over 2}\sum_{\beta}V(\vec x_\alpha,\vec x_\beta)).
\ee
$V(
\vec x_\alpha, \vec x_\beta)$ 
is the expectation value of the (density dependent) 2 body potential. 
The variation of the Lagrange function gives Euler Lagrange equations for
each of the 6 parameters
\be 
\dot{\vec p_{\alpha}} = - \vec \nabla_{\vec x_a} \sum_\beta V(
\vec x_\alpha, \vec x_\beta)
\ee
and                  
\be 
\dot{\vec x_{\alpha}} =  \vec p_\alpha/m.
\ee
With these two equations one has reduced the problem of solving a n - body
Schr\"odinger equation to that of solving 6 n ordinary differential equations. 
In reality V is a parametrization of the real part of the
Br\"uckner G- matrix. The imaginary part is approximated by the
measured cross section. For details we refer to ref.\cite{aichelin91}.  
Hence in QMD the centroids of the Gaussians in momentum and coordinate
space are the only quantities which change in time. The form of the
wave function around the centroids is fixed. This is a consequence
of the ansatz ( eq. 19). 

From the test wave function eq.(19) we calculate the 
Wigner density of a pair of particles
\be
D_{QMD}(\vec x_1 ,\vec p_1,\vec x_2,\vec p_2) = {1\over \pi^6}
e^{-(\vec P -\vec K(t))^2L/4-(\vec X -\vec R(t))^24/L}
e^{-(\vec p -\vec k(t))^2L-(\vec x -\vec r(t))^2/L}.
\ee
Inserting this Wigner density in eq. 5  one obtains after integration over 
the pair's center of mass momentum 
\be
\int{d^2W \over d\vec p_1 d\vec p_2} d^3P
= ({L\over \pi})^{3/2} (e^{-(\vec p -\vec k(t))^2L} \pm e^{- p^2L -
k(t)^2L} \cos{2\vec p \vec r})
\ee
where t is the (assumed common) freeze out time.
This is the probability to find two particles with a  relative 
momentum $\vec p$, which have been emitted from two classical sources at a
relative distance of $\vec r$ and  a relative momentum of $\vec k$.

\section{One Body Transport Theories}

In order to derive the equation for the time evolution of the one--body 
Wigner density of a particle moving in a selfconsistent potential $V(\rb)$ 
we start from that for the one body density matrix $\rho_1=|\psi_1><\psi_1|$
\be
\dot \rho_1 = -i [H,\rho_1].
\ee
Applying to this equation the Wigner transformation for an operator O
\be
O_W (\rb,\pb) = {1\over (2\pi)^3} \int d^3 y e^{i\pb\vec y} 
<\rb - {\vec y\over 2}|O|\rb + {\vec y\over 2}>
\ee
one obtains the differential equation
\be
({\partial \over \partial t} + {\pb_1 \over m}\vec \nabla_{\vec x_1}) D(\x,\p,t)
=\int d^3p_1'K_1(\pb_1-\pp_1,\rb_1)D(\rb_1,\pp_1,t)
\ee
D being the Wigner density of the one body density operator and $K_1$ 
is defined as
\be
K_1(\pb_1-\pp_1,\rb)=
{1\over i\hbar}
\int {d^3y \over (2\pi\hbar)^3}e^{-i(\pb_1-\pp_1)\vec y/\hbar} 
(V(\rb+\vec y/2)- V(\rb-\vec y/2)).
\ee
We have restored $\hbar$ here for reasons which will soon become obvious.
One can expand the integration kernel around $x_1$ 
\be
K_1(\pb_1-\pp_1,\rb_1)=
{2\over \hbar} \sin({\hbar\vec \nabla_{\vec x_1}\vec \nabla_{\vec p_1}\over 2})
V(\rb_1)\delta(\pb_1-\pp_1).
\ee
We see that $ K_1 $ can be viewed as a series with the expansion coefficient
$\hbar\vec \nabla_{\vec x_1}\vec \nabla_{\vec p_1}$. Hence in the limit that the expansion can be
terminated after the first term the Schr\"odinger equation in its
Wigner representation is equivalent to the classical Vlasov
equation :\par

\be
({\partial \over \partial t} + {\pb_1 \over m}\vec \nabla_{\vec x_1}) D(\x,\p,t)
= (\vec \nabla_{\vec x_1} V(\rb_1))\vec \nabla_{\vec p_1} D(\x,\p,t)
\ee

The Vlasov equation describes the time evolution of the phase space density
of particles which move on classical orbits specified by the Hamilton
equations ${\partial \x \over \partial t} = {\p \over m}$ and $
{\partial \p \over \partial  t}
 = - \vec \nabla_{\vec x_1} V $. As in QMD V presents the real part of the 
Br\"uckner G- matrix and the imaginary part is added as a cross section. \par

There are two approaches to solve the above equation. Either
one solves the differential equation directly or one creates
a swarm of test particles which are subject to the Hamilton equations
and fulfil the initial condition $D(\x,\p,t_0)$. One propagates
this swarm  with help of the Hamilton equations until a time t and then constructs the Wigner density
$D(\x,\p,t)$ by coarse graining. The latter solution method is called test
particle method and is employed in the BUU, VUU and LV approaches.

When calculating the observables, i.e. the expectation values
of operators, the transition from the first to the second method corresponds 
to the replacement of an analytical integration by a Monte Carlo procedure.
Using the swarm of test particles the analytical solution 
\begin{equation}
\langle O(t) \rangle = \int D(\x,\p,t) O(\x,\p) \, d^3x_1\, d^3 p_1
\end{equation}
is replaced by the corresponding Monte Carlo type integral
\begin{equation}
\langle O(t) \rangle = \frac{1}{N} \sum_{i=1}^{N}
O(\vec r_i(t),\vec k_i(t))
\end{equation}
where the $\vec r_i(t)$ and $\vec k_i(t)$ are the phase space coordinates of
the N test particles propagated with the Hamilton equations. As said, they are
distributed like $D(\x,\p,t)$. According to the theory of the Monte Carlo 
integration both integration procedures yield the same
result in the limit of an infinite number of test particles. In practice one
has to verify that the results do not depend on this number. Usually
100 test particles per physical nucleon in the system are considered as
sufficient. It is very important to realize that these test particles have
nothing to do with physical nucleons. They serve only as a representation
of the one body Wigner density $D(\x,\p,t)$. All
observables which require more than its knowledge are beyond the scope
of applicability of these theories. Hence the possibility to extract source
radii and hence correlation functions from the one body theories requires: 

\begin{itemize}
\item
The smoothness assumption is valid
\item
$D(\x,\p,\rb_2,\pb_2,t) \approx D(\x,\p,t)D(\rb_2,\pb_2,t)$
\end{itemize}
They are a consequence of the impossibility to create two body
Wigner densities or Wigner densities of two body observables
like $\vec p_1+\vec p_2$ from the swarm of test particles defined as above.
If both approximation were valid the correlation function is given by 
\bea
{d^2W \over d\vec p_1 d\vec p_2} &= \int d^3x_1d^3x_2
D(\vec x_1,\vec p_1)D(\vec x_2,\vec p_2)
(1 \pm cos(\vec p_1-\vec p_2)(\vec x_1-\vec x_2)) \nonumber \\
& = {1 \over N(N-1)}\sum_{i\neq j} (1\pm cos(\vec p_i(t_0)-\vec p_j(t_0))(\vec x_i(t_0)-\vec x_j
(t_0)))
\eea
$t_0$ is the (assumed common) freeze out time.
The second approximation, the absence of two body correlations, 
is hard to control. The importance of many particle
correlations for the fragment formation has been discussed in \cite{go}
but its relevance for the proton or pion emission has not yet 
been investigated. 

\section{Results for a given source distribution}
To interpret the different results given above it is useful to apply them
to a situation where the source is known.
To keep the things simple we assume a completely chaotic source without 
any correlation between coordinate and momentum space:
\be
S(\vec k_1,\vec r_1) = ({B\over A\pi^2})^{3/2}e^{-k_1^2B/2 -r_1^22/A}.
\ee
We start out from the Wigner density (eq.26) for a pair of particles as given
in the QMD simulation. Averaging over the Gaussian source distribution we 
obtain for the correlation function as in eq.5
\be
C(\vec p) ={\int {d^2d W \over d\vec p_1 d\vec p_2} d^3PS(\vec k,\vec r)d^3kd^3r \over
\int { dW \over d\vec p_1 }{dW \over d\vec p_2}d^3PS(\vec k,\vec r)d^3kd^3r} =   
1 \pm e^{-p^2(L+A-{LB\over L+B})}
\ee
where $S(\vec k, \vec r) = \int S(\vec k_1,\vec r_1)\cdot 
S(\vec k_2,\vec r_2) d^3K d^3R$.
If we apply the smoothness assumption (eq.6) we obtain
\be
C^{SA}(\vec p) = {\int {d^2d W^{SA} \over d\vec p_1 d\vec p_2} d^3P
S(\vec k,\vec r) d^3kd^3r
\over
\int { dW \over d\vec p_1 }{dW \over d\vec p_2}d^3PS(\vec k,\vec r)
d^3kd^3r}
 = 1 \pm e^{- p^2 (L+A)}. 
\ee
If we assume as in BUU, VUU or LV that the one body Wigner density is not given 
by Gaussians but as a sum over test particles (TP) each represented by a 
delta function in coordinate and momentum space  
\be
D^{TP}(\vec x_1,\vec p_1,t) = 
{1 \over N}\sum_{\alpha =1}^N 
\delta (\vec x_1 - \vec x_\alpha(t)) \delta (\vec p_1 - \vec p_\alpha(t))
\ee
where the $p_\alpha $'s and $x_\alpha$ 's  are distributed according to our source function we
obtain as a correlation function
\be
C^{TP}(\vec p) = 1 \pm e^{- p^2 A}. 
\ee
Defining the square of the source radii as $ {\int C(\vec p) d^3p 
\over \int C(\vec p) p^2d^3p}$ 
and comparing eqs. 38,39,41 we observe that for the same measured correlation function
$C(\vec p)$ we obtain different source radii depending
on the simulation programs and the approximations used. From a mathematical
point of view the difference between $C^{TP}(\vec p)$ and $C^{SA}(\vec p)$ is easy to 
understand.
Because the wave function used for the calculation of $C^{SA}(\vec p)$ has a width
of L, the true distribution of the source is the convolution of
the distribution of the centers given by $S(\vec r, \vec k)$ with the 
distribution of the one particle density around the centers. For
$C^{TP}(\vec p)$ one assumes that the  true source distribution is given by
$S(\vec r, \vec k)$. In order to make both quantities comparable, 
both mean square radii have to be the same and hence one has to replace
in $C^{TP}(\vec p)$ A by A'= L+A. 

However, being purely mathematical, this argument
has an essential drawback. We have started out from the approximation (eq.1)
that the source can be treated classically, and hence that the distance
between two sources is large as compared to L \cite{Podgor72}. Hence,
either the difference between A' and A is small and can be neglected. Then our
approximation is valid. Or this difference is not negligible. Then our
classical source approximation breaks down. 
That the wave function plays indeed
a nontrivial role can be seen if one compares $C^{SA}(\vec p)$  and $C(\vec p)$.
In both cases the same single particle wave function has been
employed. The result for the correlation function is, however, different.
Only if $L << A$ the difference between both is negligible. Hence
our quantitative result confirms the well known qualitative argument
that the smoothness assumption is only valid if the source can be assumed to
be classical, i.e. if the width of the wave function is small as compared
to the size of the emitting system. 
Opposite, if L is of the same order as A as in all presently employed simulation
models, the difference becomes important as we will
see below and hence the smoothness assumption will break down. Hence
we are confronted with the fact that present day simulation programs
use a value of L which neither justifies the classical treatment of the
source nor confirms the validity of the smoothness assumption. 

Nevertheless it seems that the community has agreed upon a
pragmatic point of view in pretending that at least the classical 
treatment of the source is acceptable in modeling heavy ion collisions
although a proof has not be given yet. 
Hence it may be useful to see whether under this assumption a quantitative 
prediction is possible. This includes the  answer to two questions:
To what precision we desire to measure the density and 
is the systematic error of the correlation function sufficiently small
to obtain the desired precision.

The study of the space time correlation is born out of the demand to 
measure the size of the system at the moment when the particles are emitted.
If we study nucleons of the fireball, the density has to be
in between twice and half normal nuclear
matter density, because if the expanding fireball passes the latter density,
there are no interactions anymore and hence the emission of particles
defined as the time point of the last collision has ceased. For nucleons
emitted from the spectator matter which remains at 
normal nuclear matter density one would like to know the source size.
At lower
energies the interest is to study whether the emitted nucleons come
from a compound nucleus or whether the they are emitted from a subsystem
called hot spot. Also here the density varies little around normal nuclear
matter density. Whereas in the first case an uncertainty of the density
determination of about 20\% may be tolerable, the latter two require
a precision of the determination if the source radius by about 3\% 
(and hence of the mass number of about 10\%) if
one would like to avoid that the uncertainty is already as large as the
possible variation of the size of the system under investigation.

In order to see whether this precision can be obtained we have
to calculate the values for A, L and B for the cases of interest. 
If we assume that the rms radius of the source
corresponds to the size of a nucleus at normal nuclear matter density
$R_0 = 1.2 A_M^{1/3}$ where $A_M$ is the atomic number of the nucleus
we obtain
\be
A = A_M^{2/3} [fm^2] 
\ee
i.e. A = 21.5 $fm^2$ for $A_M = 100$ and A = 34 $fm^2$ for $A_M = 200$.
Because our source emits particles according to
a Maxwell Boltzmann distribution we can relate the slope in momentum
space with the temperature of the emitting source and find that
\be
B = { 40 \over T [MeV]} [fm^2] 
\ee
In the standard versions of QMD resp. 
IQMD the parameter L has the value 4.33 and 8.66 $fm^2$, respectively.
Hence first of all we observe that L is not at all negligible as
compared to A. However, as mentioned above, accepting a classical treatment 
of the source we can correct for this. It remains to be seen whether
the smoothness assumption can be justified. Comparing the mean
square radii obtained with and without the smoothness assumption
\bea
F={R\over R^{SA}} &=& {(L+A-{LB\over L+B})\over (L+A)} \nonumber \\
&=& 1 - {LB\over (L+B)(L+A)}
\eea
we find that the smoothness assumption pretends a larger radius of the system.
The value of F ranges between .87 for small systems at low temperature (5 MeV) 
and .98 for large systems at high temperature (80 MeV). Hence for
particles emitted from a compound nucleus or from the spectator matter
the error in the determination of the mass number due to the smoothness
approximation is of the order of 20\% even if the source is completely chaotic
and of known form 
and the classical approximation of the source remains valid. For
fireball nucleons the smoothness assumption produces an error of about
4\% on the density. Of course if we were sure that we have a source of a
given temperature we could also correct for the temperature, however such
a source is not encountered in heavy ion physics where the excitation
energy and hence the temperature changes in the course of time. 

\section{Realistic Simulations}
We have seen that for the most favourable condition (chaotic source of 
known form without
any momentum space coordinate space correlation) the smoothness assumption
enlarges the apparent source size by about 20\%. If one applies now
the simulation programs to real experiments one has to inspect the 
consequences of two facts: 

1) Nature most probably does not keep the rms radius of a nuclear wave
function constant during a heavy ion reactions, QMD does. For
observables which do not depend on the width of the wave function explicitly
this may be of minor importance, the influence on observables which
depend explicitly on the width, like the correlation function, is hard
to judge since no calculations are available for a more sophisticated
treatment of the reaction as done in QMD. If the width of the wave function
has changed in the course of the reaction the difference between $C^{TP}(\vec p)$ 
and $C^{SA}(\vec p)$ cannot be corrected anymore by use of the known initial
density distribution.  

In the QMD calculations the width of the wave function  L serves two 
purposes. First it is used to have the proper one body density 
distribution when one initialize the nuclei. This is, however, a very weak
condition because with much larger widths than that actually employed
one can obtain the same one particle density distribution. Second, it
appears in the time evolution equations but only in form of the expectation
value of the potential. Thus what counts for the time evolution is the
convolution of the potential range and the width of the wave functions.
Hence one can obtain the same expectation value of the 
potential for a smaller width and a larger
potential range. Hence there is no need for an exact determination of
the width L in the QMD calculation or, vice versa, 
the success of these calculations cannot be used to determine L. 

2) The source is as simulation programs show  not at all 
chaotic and shows strong correlations between momenta and positions.

Momentum space coordinate space correlations decrease the source
size extracted from the correlation function as compared to the
geometrical size of the source. This can be easily understood
if one goes to the extreme. If the momentum is a monotonic function of
the position, two particles with a small relative momentum have to come
from the places very close in coordinates space. Thus the correlation 
function measures only that region in coordinate space from where these
particles can come. Hence the stronger the momentum space coordinate 
space correlations are the smaller is the source size measured by the
correlation function. As a consequence the value of A becomes smaller
and the importance of the width of the wave function increases. Thus
the stronger these correlations are the larger becomes the difference
between $R^{SA}$ and R (eq. 44).

Thus for realistic calculations the situation becomes worse 
as compared to a static source. For a given size of the system correlations 
make A smaller and hence increase the importance of the width of the wave function
if one compares eqs. 39 and 41 . They also
do not allow to corrected for the smoothness assumption (eqs. 38 and 39)
because the temperature is not anymore a global variable. 
The calculation of the systematic error of the value of the radius 
determined by eqs. 39 or 41  
requires more than the present models can predict, however the above arguments
show that it will be larger than that for a static source.

\section{conclusion}
We have discussed the possibility to extract source radii by
comparing the experimental results with the prediction of simulation 
programs. There is no doubt that the experimental results
indeed show momentum space correlations caused by the bosonic or fermionic
nature of the observed hadrons. These correlations carry information
about the space time structure of the reaction. The goal to
relate the observed correlation functions in momentum space with
physical parameters in coordinate space like source radii,
densities or energy densities can presently only achieved by use
of transport theories.  

None of these transport models takes the bosonic or
fermionic nature of the hadrons into account. Whereas this may be
no essential drawback for many observables it makes it impossible
to calculate the correlation function in a straight forward manner.
Every model requires the introduction of the (anti)symmetrization
of the wave function
in an ad hoc fashion in order to predict a correlation function.   

We have found that for all presently existing models, which
can be subdivided into three classes, this introduction poses problems.

Cascade models,
in which classical particles are propagated, do not allow the calculation
of a correlation function.
The quoted values of source radii are totally artificial being a 
consequence of the
employed approximation and not of physical origin.

QMD, LV, BUU and VUU models allow the calculation of a correlation function.
We find, however, that the basic approximation of the whole approach, namely
that the source can be considered as classical, is not fulfilled. Even
if it were fulfilled, the systematic error of the extracted density
introduced by the smoothness approximation is for the most favourable case 
of a chaotic source of known form up to 20\%. For realistic cases 
where space momentum space correlations are present and where we do
not know the form of the source we have
shown that the error will increase. This questions the possibility that
in nuclear physics the HBT method  allows a determination of the density
to a precision which allows to discriminate between different proposed
interaction mechanisms. 

Of course this raises the question how to proceed. 
As we have seen we are plagued with systematic errors of the order L/A.
There is first of all the open question whether the wave function of
a emitted nucleon is smaller than L. Mean field calculation yield a much
broader wave function and consequently the approximation of a classical
source cannot be justified any more.
Short range correlations, however, may distort
this wavefunction. Hence it may be justified to address the question if
there is a possibility to construct dynamical theories which can provide
a prediction for the correlation function?  Either
one can try to decrease L or to avoid the systematic errors.

The first suggestion implies a localization of the particles 
with a precision of about 1fm. This will be hardly possible.
Not only because in a nuclear environment the root mean square
radius of the wave function of the nucleon is considerably larger
than the radius of a free nucleon but also because it implies an
uncertainty of 200 MeV/c for the momentum of the nucleons 
which poses several severe technical problems for transport theories:
\begin{itemize}
\item
How to propagate particles 
in semiclassical theories whose velocity uncertainty is about 0.2c
is unknown.
\item
The sequence of collisions becomes undetermined
\item
The applied scattering cross sections have to be modified because 
the scattering partners are asymptotically not in a plane wave state.
\end{itemize}

The second suggestion implies the construction of transport theories 
which propagates at least two particle wave functions and not parameters 
of the wave function. Presently such an approach is not available.

Before a solution to these problems has been found the Hanbury Brown
Twiss effect is a very nice quantal effect. Its application in nuclear
physics to study the space time structure remains, however, premature.

Interesting discussions with Drs. Ardouin, Erazmus, Gyulassy, Heinz, Lednicky
and Werner are gratefully acknowledged. Furthermore I would like to thank
Dr. Heinz for a careful reading of the manuscript. 
  

\end{document}